\documentclass{article}
\usepackage{spconf,amsmath,graphicx}
\usepackage{enumitem}
\usepackage{cite}
\usepackage{xcolor}
\usepackage{booktabs}
\usepackage{threeparttable}
\usepackage{amsfonts,bm}
\usepackage{mathtools}
\usepackage{pifont}
\usepackage{multirow}
\usepackage{algorithm}
\usepackage{algorithmic}
\usepackage{soul}

\usepackage[bookmarks=true,breaklinks=true,colorlinks,linkcolor=blue,citecolor=blue,urlcolor=blue]{hyperref}
\usepackage{comment}
\usepackage{xspace}
\newcommand{\design}{{LSTM-QGAN}\xspace}

\title{\design: Scalable NISQ Generative Adversarial Network\vspace{-11pt}}

\name{Cheng Chu \qquad Aishwarya Hastak \qquad Fan Chen\vspace{-12pt}}
\address{Luddy School of Informatics, Computing, and Engineering, Indiana University Bloomington, USA\vspace{-10pt}}

\begin{document}
\maketitle

\begin{abstract}
Current quantum generative adversarial networks (QGANs) still struggle with practical-sized data. First, many QGANs use principal component analysis (PCA) for dimension reduction, which, as our studies reveal, can diminish the QGAN's effectiveness. Second, methods that segment inputs into smaller patches processed by multiple generators face scalability issues. 
In this work, we propose~\design, a QGAN architecture that eliminates PCA preprocessing and integrates quantum long short-term memory (QLSTM) to ensure scalable performance. 
Our experiments show that~\design~significantly enhances both performance and scalability over state-of-the-art QGAN models, with visual data improvements, reduced Fr\'echet Inception Distance scores, and reductions of 5$\times$ in qubit counts, 5$\times$ in single-qubit gates, and 12$\times$ in two-qubit gates.
\end{abstract}


\vspace{-0.12in}
\section{Introduction}
\vspace{-0.1in}

\textbf{Current QGANs}.
Recent advancements in Noisy Intermediate-Scale Quantum (NISQ) platforms~\cite{arute2019quantum, pino2021demonstration, zhong2020quantum} have catalyzed intense research on Quantum Generative Adversarial Networks (QGANs)~\cite{Lloyd2018lett, Pierre2018phyA, zoufal2019quantum, zhu2022generative, WGAN2019quantum, niu2021entangling, Stein2021qugan, chu2023iqgan, patch-gan, tsang2023hybrid}, which are well-suited to the constraints of NISQ systems, such as limited qubit counts and shallow circuit depths\cite{bharti2022noisy}.
Building on the foundational work~\cite{Lloyd2018lett} that established the theoretical superiority of QGANs over classical counterparts, 
early QGAN implementations~\cite{Pierre2018phyA, zoufal2019quantum, zhu2022generative} only focused on low-dimensional inputs like single-bit data. 
Subsequent research introduced innovations such as Wasserstein loss~\cite{WGAN2019quantum} 
and novel architectures~\cite{niu2021entangling} to improve training stability.
More recent work~\cite{Stein2021qugan, chu2023iqgan} expanded QGANs to high-dimensional data, like the 28$\times$28 MNIST dataset, by employing dimensionality reduction techniques like Principal Component Analysis (PCA).
The state-of-the-art (SOTA) PatchGAN~\cite{patch-gan} further segments inputs into smaller patches, enabling efficient processing on practical NISQ devices.

\textbf{Limitations}.
Despite recent developments, QGANs continue to face challenges in managing practical-sized data.
\underline{First}, while pre- and post-processing with PCA and inverse PCA~\cite{Stein2021qugan, chu2023iqgan} enable QGANs to handle large-dimensional data, PCA often dominates the process, diminishing the contributions of the QGANs themselves. 
\underline{Second}, although PatchGAN~\cite{patch-gan} facilitates the direct processing of practical-sized inputs through multiple small patches, its architectural limitations demand an increasing number of quantum resources as input size grows, leading to serious scalability challenges. 
For instance, generating a single MNIST image requires a prohibitively high 56 sub-quantum generators and 280 qubits.
\underline{Third}, and more concerning, our preliminary study shows a significant decline in output quality as PatchGAN scales from its original 5-qubit design~\cite{patch-gan} to 8 qubits, severely limiting its effectiveness at larger scales.

\textbf{Contributions}.
We introduce~\design, a novel architecture that eliminates the need for PCA when processing large-dimensional data. 
The design allows for the use of a constant amount of NISQ computing resources as input size increases. 
However, as additional hardware resources become available, the architecture scales efficiently, ensuring consistent and reliable performance.
Our contributions include:

\begin{itemize}[leftmargin=*, topsep=-2pt, partopsep=-2pt, itemsep=-2pt]
\item \textbf{Preliminary Analysis}.
We conduct experiments on the SOTA QGANs~\cite{Stein2021qugan, chu2023iqgan, patch-gan}, revealing previously undisclosed limitations in PCA pre-processing and model scalability.

\item \textbf{Scalable Architecture}.
We present~\design, a scalable QGAN architecture inspired by recent advances in quantum long-short memory (QLSTM)~\cite{qlstm2022quantum, di2022dawn, el2022quantum}. 
\design~eliminates the need for PCA, maintains constant NISQ resources as input size grows, and efficiently scales with increasing quantum computing resources.

\item \textbf{Enhanced Performance}.
We conduct evaluations on NISQ computers. Experimental results show that~\design~significantly enhances generative performance and improves scalability compared to SOTA QGANs.
\end{itemize}

\vspace{-0.1in}
\section{Background}
\vspace{-0.1in}

\begin{figure}[t!]
\centering
\includegraphics[width=.98\linewidth]{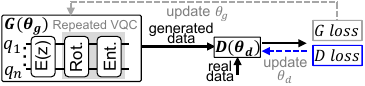}
\vspace{-0.16in}
\caption{A standard QGAN architecture.}
\label{f:background-gan}
\vspace{-0.24in}
\end{figure}

\textbf{QGAN Basics}.
Figure~\ref{f:background-gan}
illustrates a standard QGAN with two parameterized models: the Generator, $G(\theta_g)$, which generates synthetic data, and the Discriminator, $D(\theta_d)$, which evaluates the generated data against real data. 
$G$ is implemented using a quantum neural network (QNN), typically composed of a data encoder $E(\cdot)$ and repeated layers of a variational quantum circuit (VQC) with one-qubit rotations (i.e., \textit{Rot.}) and two-qubit entanglement (i.e., \textit{Ent.}). 
$D$ in SOTA QGANs~\cite{Stein2021qugan, chu2023iqgan, patch-gan} can be implemented with either classical or quantum models.
The objective is to optimize the predefined minmax loss $\mathcal{L}$, as outlined in Equation~\ref{e:objectFunc}, where $z$ represents the latent variable.
The specific loss function can be implemented using various specified functions~\cite{patch-gan, zoufal2019quantum, zhu2022generative, WGAN2019quantum, tsang2023hybrid}.
The overall goal is to enable $G$ to generate data indistinguishable from real data, while $D$ improves its ability to differentiate between them.
\vspace{-8pt}
\begin{equation}
\min_{\theta_g} \max_{\theta_d} \mathcal{L} \{ {D_{\theta_d}}({G_{\theta_g}(z)}),  {D_{\theta_d}}(x) \}
\label{e:objectFunc}
\vspace{-0.05in}
\end{equation}

\begin{figure}[t!]
\centering
\includegraphics[width=.98\linewidth]{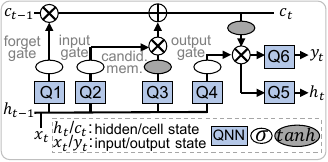}
\vspace{-0.15in}
\caption{A standard QLSTM architecture.}
\label{f:background-lstm}
\vspace{-0.25in}
\end{figure}

\textbf{SOTA QGANs}.
To manage larger-dimensional data  with limited qubits on NISQ computers, SOTA QGANs~\cite{Stein2021qugan, chu2023iqgan, patch-gan}) primarily utilize the following two techniques:

\begin{itemize}[leftmargin=*, topsep=0pt, partopsep=0pt, itemsep=-4pt]
\item \textbf{Pre- and Post-Processing}.
Several recent QGANs~\cite{Stein2021qugan, chu2023iqgan} utilize principal component analysis to reduce input dimensions (e.g., from 784 to 4 in~\cite{Stein2021qugan}) to fit within the limitations of NISQ computers with constrained qubits. 
The key steps in PCA involve: 
(1) standardizing the data to have zero mean and unit variance, and
(2) calculating the covariance matrix $\mathbf{C}$ and the matrix $\mathbf{V}_k$, which contains the top $k$ eigenvectors (principal components).
For any data matrix $\mathbf{X}$ with mean $\mu$, the data can be reduced to the top $k$ principal components by $\mathbf{Z}$=$\mathbf{X}\mathbf{V}_k$.
The reduced-dimensional data $\mathbf{Z}^*$ can then be reconstructed to approximate the original data through inverse PCA: $\mathbf{X}^*$=$\mathbf{Z}^*\mathbf{V}^\top_k$ + $\mu$.

\item \textbf{Patched Input}.
PatchGAN~\cite{patch-gan} segments the input into small regional patches and trains a dedicated sub-generator for each, capable of generating synthesized data that follows the pattern of the corresponding patch. 
This approach makes it a resource-efficient QGAN framework. 
The number of sub-generators scales with the input size; for instance, the 5-qubit design in the original work~\cite{patch-gan} requires 56 sub-generators to process the 784-pixel MNIST dataset, and doubling the input size would proportionally increase the number of sub-generators needed. 
\end{itemize}

\textbf{QLSTM}.
Long short-term memory~\cite{graves2012long} effectively captures spatiotemporal information, enabling task-specific regulation of data flow. 
Recent work~\cite{qlstm2022quantum} has introduced quantum LSTM, extended to various sequential learning tasks~\cite{di2022dawn, el2022quantum}. 
As shown in Figure~\ref{f:background-lstm}, QLSTM retains the classical LSTM gating mechanism, with the key distinction being the integration of QNNs. 
Due to page limit, we refer readers to~\cite{graves2012long, qlstm2022quantum} for detailed insights into QLSTM. 
LSTM has already been applied in classical GANs~\cite{brophy2023generative, yu2021conditional}, demonstrating enhanced generative power and reduced computational cost. 
Building on this, we aim to \textit{leverage LSTM's ability to selectively retain relevant patterns within a QGAN by training a QLSTM-based generator using different patched inputs, rather than separate sub-generators for different patches as in~\cite{patch-gan}}.


\section{Preliminary Study and Motivation}
\vspace{-0.12in}

\subsection{Preliminary Study}
\vspace{-0.08in}

\begin{figure}[t!]
\centering
\includegraphics[width=\linewidth]{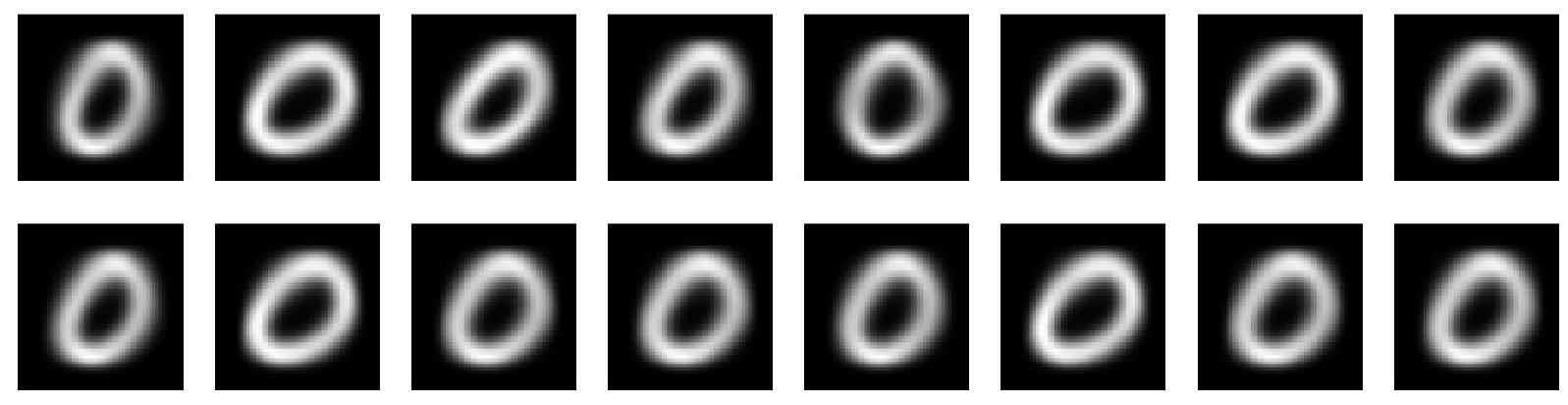}
\vspace{-0.35in}
\caption{Images reconstructed via inverse PCA using random vectors with the covariance matrix from the MNIST dataset.}
\label{f:PCA}
\vspace{-0.1in}
\end{figure}

\textbf{PCA Overshadows QGANs}.
QGANs~\cite{Stein2021qugan, chu2023iqgan} on MNIST use PCA and inverse PCA for dimensionality reduction and reconstruction.
To assess PCA's impact, we reduced 28$\times$28 MNIST images to 1$\times$2 vectors using \texttt{scikit-learn}, generating the corresponding $\mathbf{C}$, $\mathbf{V}_2$, and $\mu$. 
We then randomly generated 1$\times$2 vectors, applied inverse PCA, and present the reconstructed images in Figure~\ref{f:PCA}.
The reconstructed images closely resemble the original MNIST data and are comparable to those generated by QGANs~\cite{Stein2021qugan, chu2023iqgan}, suggesting that PCA pre- and post-processing may dominate, potentially overshadowing QGAN effectiveness. \textit{These results highlight concerns about the independent validity of QGANs when PCA is used, emphasizing the need for evaluation with unprocessed data.}

\begin{figure}[t!]
\centering
\includegraphics[width=\linewidth]{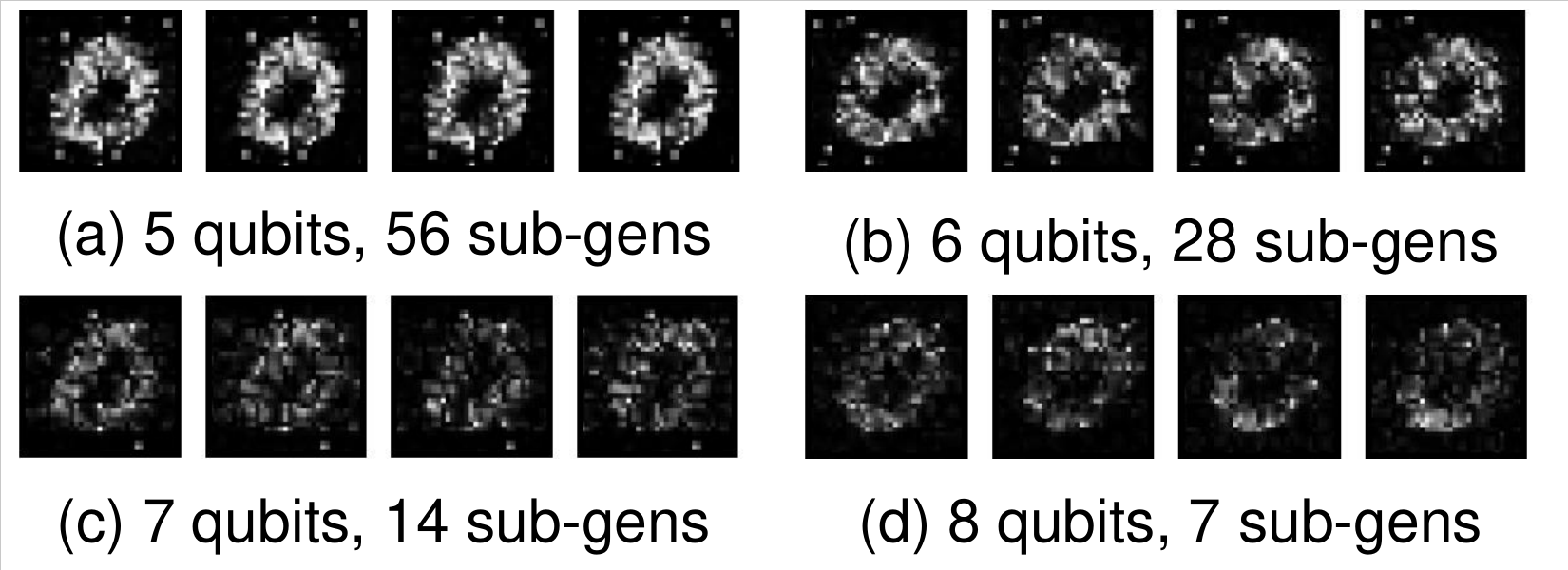}
\vspace{-0.32in}
\caption{Image quality degradation in scaling PatchGAN~\cite{patch-gan}.}
\label{f:patch_gan_explore_small}
\vspace{-0.2in}
\end{figure}

\textbf{Scalability for PatchGAN}.
PatchGAN~\cite{patch-gan} claims effectiveness with patch-based processing of high-dimensional inputs but originally reports results with only 5 qubits. 
To evaluate scalability, we increased the qubit count from 5 to 8. Since PatchGAN employs amplitude encoding and processes one patch at a time, we adjusted the number of sub-generators to cover all 784 pixels in an MNIST image.
As show in Figure~\ref{f:patch_gan_explore_small}, 
the generated images degrade rapidly with increasing qubits, with sub-figure titles indicating qubit counts and required sub-generators (sub-gens). \textit{These findings underscore PatchGAN's poor scalability, suggesting limited potential for handling larger-scale inputs even with additional qubits.}


\begin{figure*}[t!]
\centering
\includegraphics[width=1\linewidth]{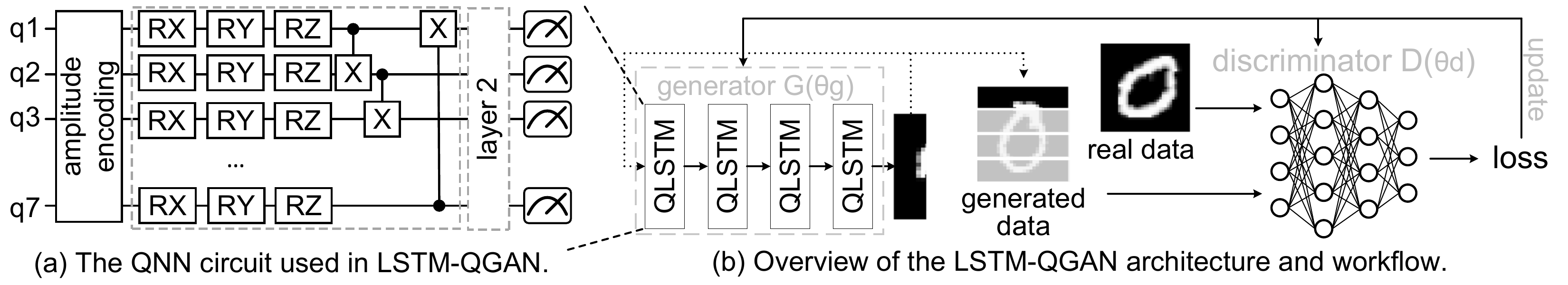}
\vspace{-0.35in}
\caption{The proposed~\design~framework.}
\label{f:overview}
\vspace{-0.25in}
\end{figure*}

\vspace{-0.15in}
\subsection{Motivation}
\vspace{-0.05in}
Our preliminary results highlight the critical need for a QGAN model capable of directly processing real-world data without PCA preprocessing, as well as a more scalable architecture to overcome the limitations of existing QGANs. Motivated by these findings, we are exploring the integration of patched inputs inspired by PatchGAN~\cite{patch-gan} to enable direct input processing without PCA. Specifically, we are investigating a scalable QGAN framework that leverages QLSTM as the generator's backbone, utilizing QLSTM's ability to capture spatiotemporal information across patches with a single generator, rather than separate sub-generators as in~\cite{patch-gan}. Additionally, we are reengineering the quantum circuit ansatz within the QLSTM structure to improve hardware efficiency, fully addressing the NISQ constraints overlooked in previous QLSTM studies like~\cite{qlstm2022quantum}.

\vspace{-0.15in}
\section{\design}
\vspace{-0.12in}

\subsection{Overall Architecture}
\label{subsec:overview}
\vspace{-0.1in}

As illustrated in Figure~\ref{f:overview}(b), \design~utilizes QLSTM at the core of the generator to enhance scalability and resource efficiency. 
Like PatchGAN~\cite{patch-gan}, the discriminator in~\design~can be implemented using either a classical or quantum neural network, depending on the available quantum computing resources. 
The following outlines the key components and configurations in~\design.

\begin{itemize}[leftmargin=*, nosep, topsep=0pt, partopsep=0pt]
\item \textbf{Patch Inputs without PCA}.
In line with~\cite{patch-gan, tsang2023hybrid}, \design~processes patched inputs to generate corresponding output patches, which are then recombined into a complete output. Unlike~\cite{Stein2021qugan, chu2023iqgan}, \design~eliminates the need for PCA and inverse PCA, processing the original data directly. This introduces a trade-off between resources (i.e., qubit number $N$) and processing latency (i.e., steps $T$). 
With an $N$-qubit implementation, \design~generates $2^N$ measured probabilities at each step as the output vector for each synthetic patched output. These vectors are then compared to the real patched input data in the discriminator. The total number of steps, $T$, is determined by $D/2^N$, where $D$ represents the size of the real data. 

\item \textbf{Scalable QGAN with LSTM}.
The generator in~\design~consists of four QLSTM cells, as shown in Figure~\ref{f:overview}(b). 
Normally distributed noise $z$ is input to the generator, producing the initial sub-image $G_{\theta_g}(z)$. The discriminator evaluates both synthetic and real input patches, computing the loss $\mathcal{L}$.
Unlike PatchGAN~\cite{patch-gan}, which requires separate generators for each patch—drastically increasing NISQ resource overhead with input size—\design~utilizes QLSTM's ability to retain relevant patterns while discarding irrelevant information, independent of patch indices. 
To achieve this, gradients from all patches within a single input are averaged and applied to update the model parameters collectively, resulting in an image-adaptive generator that scales with increasing data dimensions while maintaining fixed resource usage.

\item \textbf{Training Optimization}.
Convergence in QGAN training is a critical challenge, significantly influenced by the choice of quantum loss function.
Within the~\design~framework, we evaluated both the conventional binary cross entropy loss~\cite{patch-gan, zoufal2019quantum, zhu2022generative} and the Wasserstein loss~\cite{WGAN2019quantum, tsang2023hybrid}.
The specific Wasserstein loss used for~\design~is detailed in Equation~\ref{e:w_loss},
where 
$\mathbb{L}_{\hat{x}}$ = $\underset{{\hat{x} \sim P_{\hat{x}}}}{\mathbb{E}} [ \left ( \left \| \nabla _{\hat{x}} {D_{\theta_d}}(\hat{x})  \right \|_{2} -1  \right ) ^{2}]$,
$P_{r}$ and $P_{g}$ represent the real data (i.e., $x$) and 
generated data (i.e., $\tilde{x}$$\in$${D_{\theta_d}}({G_{\theta_g}(z)})$) distributions, respectively.
The distribution
$P_{\hat{x}}$ is uniformly sampled between $P_{r}$ and $P_{g}$, and
$\lambda$ is a constant.
Experimental results on the impact of QGAN loss functions are discussed in Section~~\ref{sec:exp}.

\vspace{-0.25in}
\begin{equation}
\min_{\theta_g} \max_{\theta_d} \underset{{x\sim P_{r}}}{\mathbb{E}} \left [ {D_{\theta_d}}(x)  \right ] - \underset{{\tilde{x} \sim P_{g}}}{\mathbb{E}} \left [ {D_{\theta_d}}(\tilde{x})  \right ] - 
\lambda{\mathbb{L}_{\hat{x}}}
\label{e:w_loss}
\vspace{-0.15in}
\end{equation}
\end{itemize}

\begin{table}[t!]
\vspace{-0.1in}
\caption{Design cost comparison.}
\label{tab:cost}
\vspace{0.02in}
\footnotesize
\centering
\setlength{\tabcolsep}{4pt}
\begin{tabular}{|l|c|c|}\hline
& \textbf{PatchGAN}~\cite{patch-gan} & \textbf{\design} ($\Delta$) \\ \hline\hline
\textbf{Qubits per QNN}      & 5         & 7          \\ \hline
\textbf{Number of QNNs}      & 56        & 8          \\ \hline\hline
\textbf{Total Number of Qubits}        & 280       & 56 (\textcolor{red}{5$\times$}$\mathbf{\downarrow}$)        \\ \hline
\textbf{Total Number of 1QG}       & 1680      & 336 (\textcolor{red}{5$\times$}$\mathbf{\downarrow}$)       \\ \hline
\textbf{Total Number of 2QG}       & 1344      & 112 (\textcolor{red}{12$\times$}$\mathbf{\downarrow}$)       \\ \hline
\end{tabular}
\vspace{-0.2in}
\end{table}

\vspace{-0.1in}
\subsection{NISQ Implementation}
\vspace{-0.08in}

\design~offers flexibility in implementing $G$ and $D$.
For fair comparison, $D$ is implemented as a classical neural network, as in PatchGAN~\cite{patch-gan}.
In the QLSTM cells for $G$, we employed a hardware-efficient ansatz inspired by recent QNNs~\cite{Sukin2019_vqcc14, chu2022qmlp, PatelST22vqcdate}, instead of the generic circuit from~\cite{qlstm2022quantum}.
Figure~\ref{f:overview}(a) shows the QNN circuit, which utilizes seven qubits.
Each VAC block includes \texttt{Rx}, \texttt{RY}, and \texttt{RZ} layers, followed by 2-qubit \texttt{CX} entanglement layer, with the VQC layers repeated twice.
The measurement layer converts the quantum state into classical vectors. 
Although the gate count matches that in~\cite{qlstm2022quantum}, our circuit uses native gates, while the \texttt{R}($\alpha$, $\beta$, $\gamma$) gate in~\cite{qlstm2022quantum} requires synthesis into multiple native gates.

\begin{figure*}[t!]
\centering
\includegraphics[width=\linewidth]{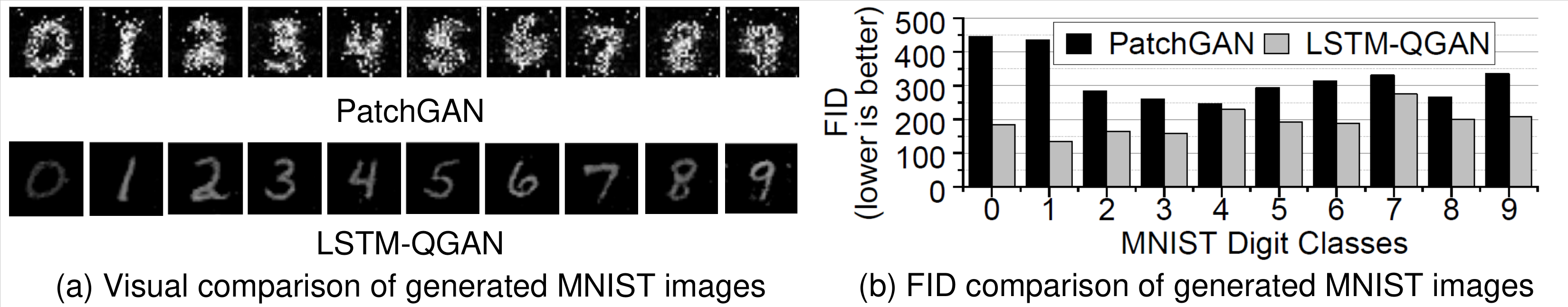}
\vspace{-0.35in}
\caption{Comparisons between images generated by PatchGAN and LSTM-QGAN.}
\label{f:LSTM-QGAN_image}
\vspace{-0.2in}
\end{figure*}

\textbf{Design Overhead}.
Table~\ref{tab:cost} compares the hardware resources required by PatchGAN and~\design~for the MNIST dataset. Due to architectural differences, a QNN in PatchGAN refers to the quantum generator used for each input patch, while in~\design, it refers to the quantum module within the QLSTM. 
The last three rows of Table~\ref{tab:cost} highlight that \design~achieves a significant reduction:
a 5$\times$ decrease in qubit counts, a 5$\times$ decrease in one-qubit gates (1QG), and a 12$\times$ decrease in two-qubit gates (2QG).

\vspace{-0.15in}
\section{Experiments and Results}
\label{sec:exp}
\vspace{-0.1in}

\subsection{Experimental Setup}
\vspace{-0.07in}

\textbf{Schemes and Benchmarks}.
We compare~\design~with PatchGAN~\cite{patch-gan} using the MNIST dataset, which consists of 28$\times$28 grayscale images of handwritten digits 0$\sim$9.
PatchGAN is implemented according to its original design~\cite{patch-gan}, utilizing 5 qubits and 56 sub-generators. Each sub-generator produces a 14-pixel patch, and together, the 56 sub-generators generate the entire 784-pixel MNIST image.
For~\design, we implement the generator with two QLSTM layers, each containing 4 QNNs with 7 qubits. At each time step, the LSTM-QGAN generates a 196-pixel patch, requiring 4 time steps to produce a complete MNIST image.

\textbf{Simulation}. 
All QGANs are implemented with the PennyLane and Torchquantum libraries.
PatchGAN and~\design~are trained using the ADAM optimizer with a 2e-4 learning rate, a 128 batch size, and 1000 epochs. Quantum circuits are run on the NISQ \texttt{IBM\_kyoto} computer~\cite{IBMQ}.

\textbf{Evaluation metrics}.
We evaluate the generated images using both qualitative (e.g., visual inspection) and quantitative methods. For quantitative assessment, we employ the Fr\'echet Inception Distance (FID), a widely recognized metric for measuring image similarity in GANs~\cite{kynkaanniemi2022role}. A lower FID score indicates a closer feature distance between real and generated images, signifying higher quality. In our experiments, we randomly select 500 real images and 500 generated images for comparison.

\vspace{-0.2in}
\subsection{Results and Analysis}
\vspace{-0.05in}

\textbf{Comparison of Image Visual Quality}.  
Figure~\ref{f:LSTM-QGAN_image}(a)
presents a visual comparison between the images generated by PatchGAN and~\design. 
PatchGAN demonstrates limited generation capabilities, as the outlines of the digits (0$\sim$9) are only vaguely identifiable, with noticeable white noise in the background. 
Additionally, the clarity of more complex digits, such as 4, 5, and 9, is particularly low, further highlighting its deficiencies.
In contrast, \design~demonstrates superior image generation, producing sharper and more distinct digits with minimal noise, underscoring its enhanced capability in generating high-quality images.

\textbf{Comparison of Image FID Scores}.
Figure~\ref{f:LSTM-QGAN_image}(b)
compares the FID scores of images generated by PatchGAN and~\design across different digit classes. 
The FID scores vary between the two models depending on the complexity and distinctiveness of each digit. 
Overall, \design~achieves lower FID scores than PatchGAN, indicating higher quality in the generated images. 
Specifically, PatchGAN shows significant variability, with its highest FID score at 445.22 (class 0) and its lowest at 246.56 (class 4). 
In contrast, \design~consistently outperforms PatchGAN, with its highest FID score at 275.58 (class 7) and its lowest at 134.31 (class 1). 
On average, PatchGAN's FID score is 318.02, while~\design~achieves a significantly lower average FID score of 193.28.

\begin{figure}[t!]
\centering
\includegraphics[width=\linewidth]{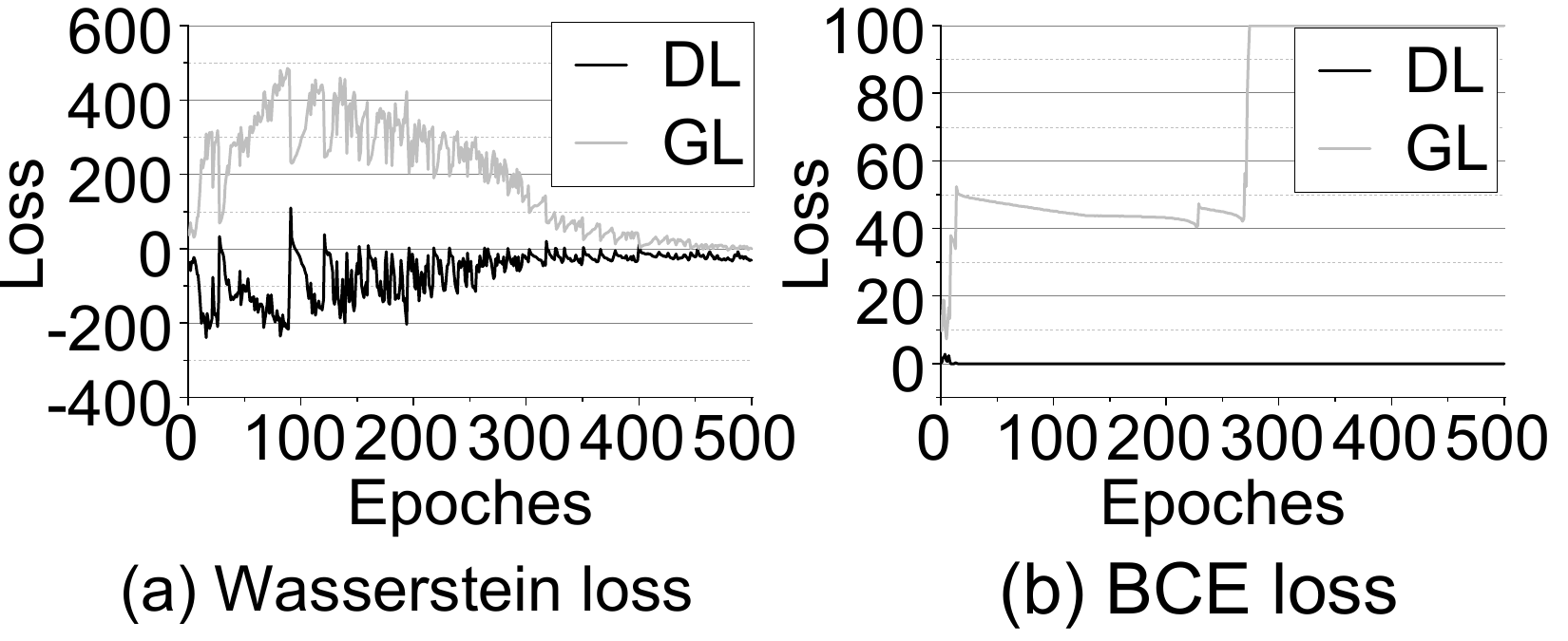}
\vspace{-0.3in}
\caption{Impact of loss functions on convergence.}
\label{f:loss}
\vspace{-0.25in}
\end{figure}

\textbf{Impact of Loss Function}.
Figure~\ref{f:loss}
illustrates the impact of Wasserstein loss and binary cross-entropy (BCE) loss on the training convergence of~\design, comparing both generator loss (i.e., GL) and discriminator loss (i.e., DL). 
With Wasserstein loss, the DL initially decreases while GL increases, ultimately leading to convergence as training progresses. 
Conversely, with BCE loss, GL rapidly increases after several training cycles and stabilizes around 100, while DL drops sharply—indicating mode collapse, a known issue in GAN training. 
Although~\design~with BCE loss can stabilize, achieving full convergence may require more sophisticated techniques. 
In contrast, Wasserstein loss offers greater training stability, resulting in smoother convergence.

\vspace{-0.1in}
\section{Conclusion}
\vspace{-0.1in}
This work presents~\design, a quantum generative adversarial network (QGAN) architecture that overcomes key limitations in existing models. By eliminating reliance on principal component analysis (PCA) and integrating quantum long short-term memory (QLSTM), \design~achieves scalable performance with efficient resource use. As the first QGAN to incorporate QLSTM, this approach represents a significant advancement likely to inspire further research.

\vspace{-0.2in}
\section*{Acknowledgment}
\vspace{-0.1in}
We thank the anonymous reviewers for their constructive and insightful comments.
This work was supported in part by NSF OAC-2417589 and NSF CNS-2143120. 
We thank the IBM Quantum Researcher \& Educators Program for their support of Quantum Computing Credits.

\vspace{-6pt}
\bibliographystyle{IEEEbib}
\bibliography{reference}

\end{document}